\documentclass[%
preprint,
a4paper,
showpacs, 
amsmath,amssymb,
prd,
longbibliography,
lengthcheck,%
]{revtex4-1}

\usepackage{graphicx}
\usepackage{dcolumn}
\usepackage{bm}
\usepackage{hyperref}
\usepackage{natbib}

\newcommand{\eq}{\begin{equation}}
\newcommand{\eqend}{\end{equation}}
\newcommand{\eqar}{\begin{eqnarray}}
\newcommand{\eqarend}{\end{eqnarray}}
\newcommand{\ovl}{\overline}
\newcommand{\A}{\alpha}
\newcommand{\B}{\beta}

 \addtocounter{section}{0}
 \makeatletter \@addtoreset{equation}{section}
 \makeatother

\begin{document}


\title{Supersymmetric QCD in the Noncommutative Geometry}


\author{Satoshi Ishihara}
  \email{satoshi@yukawa.kyoto-u.ac.jp}
\author{Hironobu Kataoka}
  \email{hrkataoka@hyogo-c.ed.jp}
\author{Atsuko Matsukawa}
  \email{matsukawa@cap.ocn.ne.jp}
\author{Hikaru Sato}
  \email{hikaru\underline{ }sato@gakushikai.jp}
\affiliation{Department of Physics, Hyogo University of Education \\
Shimokume, Kato-shi, Hyogo 673-1494, Japan}
\author{Masafumi Shimojo}
  \email{shimo0@ei.fukui-nct.ac.jp}
\affiliation{Department of Electronics and Information Technology \\
Fukui National College of Technology \\
Geshi-cho, Sabae-shi, Fukui 916-8507, Japan}


\date{December 9, 2011}

\begin{abstract}
Introduction of supersymmetry into the noncommutative geometry is investigated. We propose a new Dirac operator which plays the role of  the metric over the extended algebra of chiral and antichiral supermultiplets and  is invariant under the supersymmetry transformations.   Inner automorphisms for the algebra  generate vector  supermultiplets as an internal fluctuation of the metric. We show that the supersymmetric QCD action for these supermultiplets is correctly given by the spectral action principle.
\end{abstract}

\pacs{11.10.Nx, 11.30.Pb, 12.10.-g, 12.60.Jv}

\maketitle

\section{Introduction}
%

An application of noncommutative geometry (NCG) to particle physics was initiated by Connes \cite{connes0} and subsequent works followed~\cite{connes1,connes2,connes3,connes9}.  This approach made possible full geometric description of the interaction of elementary particles \cite{[{For a review, see, }] schucker,connes11}. Following the scheme of NCG, the noncommutative geometric standard model was studied by various authors \cite{lott,kastler,CPmartin,jureit}. A remarkable feature of this geometric approach is that both the gauge fields and the Higgs fields are introduced by internal fluctuations of the metric of NCG.  In the traditional spontaneously broken gauge theory,  the Higgs Lagrangian contains many free parameters so that they spoil the predictive power of the model. In the NCG models, the Higgs field emerges on the same footing as the gauge field and the Higgs coupling constants are related to the gauge coupling constants.

The NCG standard model, extended to include neutrino masses, was constructed \cite{connes4,connes5}.  Although the standard model of elementary particles provides a remarkably successful description of presently known phenomena, there are some difficulties and they suggest the existence of  new physics beyond the standard model. One of them is  that the gauge coupling unification in the renormalization group equation is not viable phenomenologically in the framework of the minimal Standard Model. In addition, there is the infamous "hierarchy problem", which is that the Higgs squared mass parameter $m_H^2$ receives enormous quantum corrections from the virtual effects of every particle that couples to the Higgs field. These difficulties also exist in the NCG standard model.

It is known that these shortcomings are remedied by introducing supersymmetry into the standard model \cite{[{For a review, see, }] martin}. In order for the NCG standard model to be phenomenologically viable, it is quite desirable to incorporate supersymmetry into the model. The purpose of this paper is to investigate how to introduce supersymmetry in NCG. Following our prescription of supersymmetric NCG we shall show that the supersymmetric QCD action is correctly derived.
 
The fundamental aspect of supersymmetry in NCG was discussed by Connes \cite{connes0}. Subsequently the early attempts to incorporate supersymmetry in NCG were made by several authors \cite{hussain1,hussain2,chamseddine,kalau}.
Recently the supersymmetric QCD was derived in \cite{thijs} on the basis of the spectral action principle \cite{connes8}.  In their works the superpartners of the QCD-particles were added in such a way that fermions are elements of the Hilbert space while bosons arise as inner fluctuations of a Dirac operator.  It seems rather unnatural, however, that the fermions and bosons in the same  supermultiplet  have the different origin.  
 
Our approach is different from those previous works as outlined below.
The basic element of NCG consists of an involutive algebra  of operators   and of a selfadjoint  Dirac operator in the Hilbert space. These ingredients of NCG are formulated in the Euclidean space-time.
In order to incorporate supersymmetry, however, we must work in the Minkowskian space-time.   Moreover, we enlarge the Hilbert space and the algebra in such a way that they include both fermions and bosons which form a supermultiplet.  In this situation we construct a new generalized supersymmetric Dirac operator, and show that the supersymmetric action is derived correctly from the generalized  supersymmetric Dirac operator by the use of the spectral action formula. In order to check the axioms of NCG \cite{connes7}, we change the signature from Minkowskian to Euclidean by the Wick rotation, since the axioms of NCG have been stated in the Euclidean space-time. 

Supersymmetric extension of the gauge symmetry is derived from inner automorphisms for the algebra of supermultiplets. We shall see that the effect of inner automorphisms on the metric gives rise to internal fluctuations and replace the supersymmetric Dirac operator by the modified one. This mechanism generates exactly the gauge supermultiplets. Then we can take a step toward the construction of the supersymmetric QCD in our geometric approach. It is shown that the supersymmetric QCD action is derived on the basis of the spectral action principle.
For the sake of simplicity we consider in this paper the flat space without the gravity couplings.

The present paper is organized as follows: 
 In section 2, we consider a functional space $\mathcal{H}_M$ which consists of spinor and scalar functions on the Minkowskian space-time. The supersymmetry transformation is imposed on these functions. We also introduce an algebra $\mathcal{A}_M$, whose elements are chiral and antichiral supermultiplets under the supersymmetry transformation.  Then we define the supersymmetric operator $\mathcal{D}_M$ in $\mathcal{H}_M$, which turns out to be the generalized Dirac operator if we change the signature from the Minkowskian to the Euclidean one. 

In section 3, we show that the $(\mathcal{A}_M, \mathcal{H}_M, \mathcal{D}_M) $ is the spectral triple of the Euclidean NCG.  In the Euclidean signature,  $\mathcal{H}_M$ turns out to be the Hilbert space and $\mathcal{D}_M$ is the Dirac operator with the compact resolvent. We can define the $Z/2$ grading of $\mathcal{H}_M$, which corresponds to the chirality of the supersymmetry transformation.

In section 4, the vector supermultiplet is introduced as the internal fluctuation on the metric with respect to the supersymmetric Dirac operator $\mathcal{D}_M$.  The vector fields as well as their superpartners are constructed as the bilinear form of the functions in the chiral and antichiral supermultiplets. We derive the modified  supersymmetric Dirac operator which incorporates the vector supermultiplet.     

In section 5, internal degrees of freedom are incorporated by introducing finite space with the spectral triple  $(\mathcal{H}_F, \mathcal{A}_F, \mathcal{D}_F )$. In this paper we consider quarks and color degrees of freedom by letting $\mathcal{A}_F$ be the algebra of $3 \times 3$ complex matrices and $\mathcal{H}_F$ be the Hilbert space with the basis of internal degrees of freedom of quarks.  

In section 6, we calculate the supersymmetric QCD action by applying the spectral action principle. Since the supersymmetric Dirac operator $\mathcal{D}_M$ is expressed in the Minkowskian signature, we change the signature to the Euclidean one in order to calculate the spectral action using the formula for the operator of Laplace type.  After the Wick rotation back to Minkowskian signature, we can finally obtain the supersymmetric  QCD action integral.

Finally, in section 7 
 we give our conclusions and outlook. 
%
\section{Supersymmetry and noncommutative geometry}
The basic element of NCG consists of an involutive algebra $\mathcal{A}$ of operators in Hilbert space $\mathcal{H}$ and of a selfadjoint unbounded operator $\mathcal{D}$ called the Dirac operator  in $\mathcal{H}$. A set of $(\mathcal{A},\mathcal{H},\mathcal{D})$ is named a spectral triple or a $K$-cycle \cite{connes0}. If we consider a flat space, the space-time manifold of NCG  is Euclidean.   
In order to introduce supersymmetry, however, we must work in the Minkowskian space-time $M$. So, our strategy is the following:  First we consider $(\mathcal{A}_M,\mathcal{H}_M,\mathcal{D}_M)$ on the Minkowskian space-time manifold $M$, and introduce there the supersymmetry degrees of freedom.  Then we relate it to the spectral triple of the Euclidean NCG by passing from the Minkowskian signature
$
g^{\mu\nu} = (-1,1,1,1)
$
 to the Euclidean signature 
$
\eta^{\mu\nu} = (1,1,1,1)
$
 by the Wick rotation, $t \rightarrow it$.  
When we calculate the spectral action, we work in the Euclidean signature  and then come back to the the Minkowskian signature to obtain the physical results. 

In order to introduce supersymmetry we prepare the functional space $\mathcal{H}_M$ with spinor and scalar functions of $C^\infty (M)$. 
We consider the subsets $\mathcal{H}_+$ and $\mathcal{H}_-$ of  $\mathcal{H}_M$.  The elements of  $\mathcal{H}_+$ are denoted by
\eq
(\Psi_{+})_i = \left( \varphi_+(x), \psi_{+\A}(x), F_+(x) \right),
\label{eq2.1} 
\eqend
in the vector notation, $i = 1,2,3$. Here  $\varphi_+(x)$ and $ F_+(x)$ are complex scalar functions with mass dimension 1 and 2, respectively, and $\psi_{+\A}(x), \A = 1,2$ are the Weyl spinors on the space-time manifold $M$ which have mass dimension $\frac{3}{2}$. 

The functions in Eq.(\ref{eq2.1}) are assumed to obey the following supersymmetry transformation of the chiral supermultiplet:
\eq
\left\{
\begin{array}{lcl}
\delta_\xi \varphi_+ &=& \sqrt{2} \xi^{\A} \psi_{+\A}, \\ 
\delta_\xi \psi_{+\A} &=& i \sqrt{2} \sigma^\mu_{\A\dot{\A}} \bar{\xi}^{\dot{\A}} \partial_\mu \varphi_+ + \sqrt{2} \xi_\A F_+ ,
\\
\delta_\xi F_+ &=& i \sqrt{2} \, \bar{\xi}_{\dot{\A}} \ovl{\sigma}^{\mu\dot{\A}\A} \partial_\mu \psi_{+\A}.
\end{array} 
\right.
\label{eq2.2}
\eqend
The element of $\mathcal{H}_-$ is denoted by the vector with the index $\bar{i} = 1,2,3,$ and is the following antichiral supermultiplet,
\eq
({\Psi}_{-})_{\bar{i}} = \left( \varphi_-(x),  {\psi}_{-}^{\dot{\A}}(x),  F_-(x) \right) .
\label{eq2.3}
\eqend
The functions in $\Psi_-$ obey the antichiral supersymmetry transformation:
\eq
\left\{
\begin{array}{lcl}
\delta_{{\xi}} \varphi_- &=& \sqrt{2} \bar{\xi}_{\dot{\A}} \, {\psi}_-^{\dot{\A}}, \\ 
\delta_{{\xi}} {{\psi}}^{\dot{\A}}_- &=& i \sqrt{2}  \ovl{\sigma}^{\mu\dot{\A}\A} \,{\xi}_\A \partial_\mu \varphi_- + \sqrt{2}\, \bar{\xi}^{\dot{\A}} F_-, \\
\delta_{{\xi}} F_- &=& i \sqrt{2} \,{\xi}^\A \sigma^\mu_{\A\dot{\A}} \partial_\mu {\psi}_-^{\dot{\A}}.
\end{array} \right.
\label{eq2.4}
\eqend

Weyl spinors with undotted indices ${\A}, (\A = 1,2)$ transform as the $(\frac{1}{2},0)$ representation of the Lorentz group, $SL(2,C)$, while those with dotted indices $\dot{\A}, (\dot{\A} = 1, 2)$ transform as the $(0, \frac{1}{2})$ representation. The indices $\A$ are raised and lowered with the antisymmetric tensors $\varepsilon^{\A\B}$ and $\varepsilon_{\A\B}$, where $\varepsilon^{12} = \varepsilon_{21} = 1$. The same holds for dotted indices $\dot{\A}$.  The $\varepsilon$-tensor is also used to raise the indices of the $\sigma$-matrices as $\ovl{\sigma}^{\mu\dot{\A}\A} = \varepsilon^{\dot{\A}\dot{\B}} \varepsilon^{\A\B} \sigma^\mu_{\B\dot{\B}}$.
Here and in what follows we use the notation and convention of \cite{wess}.  
We assume that the functions of chiral and antichiral supermultiplets and their derivatives are $C^\infty(M)$. 

The  algebra in the space $\mathcal{H}_M =  \mathcal{H}_+ \oplus \mathcal{H}_-$ is now defined by $\mathcal{A}_M = \mathcal{A}_+ \oplus \mathcal{A}_-$.  The element of $\mathcal{A}_+$ is given in the  following matrix form;
\eq
(u_a)_{ij} = \frac{1}{m_0} \begin{pmatrix}
\varphi_a  & 0 & 0 \\
\psi_{a\A} & \varphi_a  & 0 \\
F_a & -\psi_a^\A & \varphi_a 
\end{pmatrix} \mbox{ } \in \mbox{ } \mathcal{A}_+,
\label{eq2.5}
\eqend
where $ \varphi_a, \psi_{a\A}$ and $F_a$ form a chiral supermultiplet and they obey the supersymmetry transformation (\ref{eq2.2}). 
In Eq.(\ref{eq2.5}),  $m_0$ stands for the mass parameter which was inserted to adjust the mass dimension.
The multiplication rule among the elements in $\mathcal{A}_+$ is now given by
\begin{align}
& \hspace{1.8mm} (u_3)_{ik} =  (u_1)_{ij} (u_2)_{jk},
\label{eq2.6} \\
&\left\{
\begin{array}{lcl}
\varphi_3 &=& \varphi_1 \varphi_2 /m_0, \\
\psi_{3 \A} &=& (\psi_{1\A} \varphi_2 + \varphi_1 \psi_{2\A})/m_0, \\
F_{ 3} &=& (\varphi_1 F_{ 2} + F_{ 1} \varphi_2 - \psi^\A_1 \psi_{2\A})/m_0,
\end{array}
\right.
\label{eq2.7}
\end{align}
where $\varphi_3, \psi_{3\A}$ and $F_3$ transform again  as Eq.(\ref{eq2.2}) and belong to $\mathcal{A}_+$. 
In Eq.(\ref{eq2.6}) and in what follows,  repeated matrix indices should be understood to take summation over the matrix size.
Complex conjugate element of Eq.(\ref{eq2.5}) belongs to $\mathcal{A}_-$ and is given by
\eq
(\ovl{u}_a)_{\bar{i} \bar{j}} = \frac{1}{m_0} \begin{pmatrix}
\varphi_a^* & 0 & 0 \\
{\ovl{\psi}}_a^{\dot{\A}} & \varphi_a^* & 0 \\
F_a^* & -\ovl{\psi}_{a\dot{\A}} & \varphi_a^*
\end{pmatrix} \mbox{ } \in \mbox{ } \mathcal{A}_-,
\label{eq2.8}
\eqend
which obeys the supersymmetry transformation  for the antichiral supermultiplet given by Eq.(\ref{eq2.4}).

The algebra $\mathcal{A}_+$ is represented in the space $\mathcal{H}_+$ by the following formula,
\begin{align}
&\hspace{5mm} (\Psi'_+)_i = (u_a)_{ij} (\Psi_+)_j ,
\label{eq2.9} 
\end{align}
We have a similar relation for the antichiral sector.

The third ingredient is the generalized Dirac operator which is an operator in the space $\mathcal{H}_M$.  Since we defined $\mathcal{H}_M$ and $\mathcal{A}_M$ so as to incorporate supersymmetry, let us define the supersymmetric Dirac operator by the following matrix form:
\begin{align}
&\mathcal{D}_M = -i \begin{pmatrix}
0 & \ovl{\mathcal{D}}_{i \bar{j}} \\
{\mathcal{D}}_{\bar{i} j} & 0
\end{pmatrix},
\label{eq2.11}
\end{align}
where
\begin{align}
&\mathcal{D}_{\bar{i}{j}} = \begin{pmatrix}
0 & 0 & 1 \\
0 & i\ovl{\sigma}^\mu \partial_\mu & 0 \\
\square & 0 & 0
\end{pmatrix}, \mbox{ }
\ovl{\mathcal{D}}_{i \bar{j}} = \begin{pmatrix}
0 & 0 & 1 \\
0 & i{\sigma}^\mu \partial_\mu & 0 \\
\square & 0 & 0
\end{pmatrix},
\label{eq2.12}
\end{align}
and $\square = \partial^\mu \partial_\mu$.

The operator $\mathcal{D}_{\bar{i}{j}}$ operates on $(\Psi_+)_j$ and generates the element in $\mathcal{H}_-$ in the following way,
\begin{align}
&\hspace{5mm} (\Psi'_-)_{\bar{i}} = \frac{1}{m_0} \mathcal{D}_{\bar{i}{j}} (\Psi_+)_j,
\label{eq2.13} \\
&\left\{
\begin{array}{lcl}
\varphi'_-(x) = F_+(x)/m_0, \\
{\psi}'^{\dot{\A}}_-(x) = i \ovl{\sigma}^{\mu\dot{\A}\A} \partial_\mu \psi_{+\A} (x) /m_0, \\
F'_-(x) = \square \varphi_+(x)/m_0.
\end{array}
\right.
\label{eq2.14}
\end{align}
As a matter of fact, we can confirm that the left-hand side of Eq.(\ref{eq2.14}) transforms as Eq.(\ref{eq2.4}) for the antichiral supermultiplet by  applying Eq.(\ref{eq2.2}) on the right-hand side.  Similarly, for the operator $\ovl{\mathcal{D}}_{i \bar{j}}$ we have the following formula:
\begin{align}
&\hspace{5mm} (\Psi'_+)_{{i}} = \frac{1}{m_0} \ovl{\mathcal{D}}_{{i}\bar{j}} (\Psi_-)_{\bar{j}}.
\label{eq2.15} 
\end{align}
Let us denote the element in $\mathcal{H}_M = \mathcal{H}_+ \oplus \mathcal{H}_-$ as
\eq
\Psi = \begin{pmatrix}
\Psi_+ \\
\Psi_-
\end{pmatrix},
\label{eq2.16}
\eqend
then Eq.(\ref{eq2.13}) and Eq.(\ref{eq2.15}) are put together in the following form:
\eq
\Psi' = \frac{i}{m_0} \mathcal{D}_M \Psi .
\label{eq2.17}
\eqend
By applying the supersymmetry transformation, Eq.(\ref{eq2.2}) and Eq.(\ref{eq2.4}), we can show that
\begin{align}
&\left[ \delta_\xi, \mathcal{D}_M \right] = 0.
\label{eq2.18}
\end{align}
Thus the operator $\mathcal{D}_M$ given by Eq.(\ref{eq2.11}) is invariant under the supersymmetry transformation.

Here a brief comment is in order.  If we introduce  anticommuting parameters $\theta^\A, \bar{\theta}^{\dot{\A}}$, an element of the algebra may be expressed by a superfield. For example, Eq.(\ref{eq2.5}) corresponds to the following chiral superfield,
\eq
\Phi_a(x_+) = \varphi_a(x_+) + \sqrt{2} \theta \psi_a(x_+) + \theta\theta F_a(x_+), 
\label{eq2.19}
\eqend
where $x_+^\mu = x^\mu + \theta \sigma^\mu \bar{\theta}$.  Then the multiplication rule, Eq.(\ref{eq2.7}) is obtained since products of two chiral superfields are again chiral superfields.  
Note that the set of superfields does not form
the Hilbert space since the norm of the superfield is not defined.  This is the reason why we constructed $\mathcal{H}_M$ and $\mathcal{A}_M$ using the component fields. 
%
%
\section{The spectral triple in the framework of supersymmetry}

 We first examine that the  $(\mathcal{A}_M, \mathcal{H}_M, \mathcal{D}_M)$  introduced in the previous section  play the role of the spectral triple of NCG.  Since NCG is formulated in the framework of the Euclidean signature, we change the signature from the Minkowskian to the Euclidean one by the Wick rotation.  

The functional space $\mathcal{H}_M$ has been defined to include the scalar functions in addition to the spinor functions in such a way that they form a supermultiplet. In the Euclidean signature, the algebra of $SL(2,C)$ turns out the algebra of $SU(2) \otimes SU(2)$ under the Wick rotation. So, the Weyl spinors $\psi_{+\A}$ and ${\psi}_{-}^{\dot{\A}}$ in $\mathcal{H}_M$ are replaced by $\rho_\A$ and $\omega^{\dot{\A}}$ where they transform as $(\frac{1}{2}, 0)$ and $(0, \frac{1}{2})$ of $SU(2) \otimes SU(2)$, respectively.  Here $\A = 1,2$ is the index of the first $SU(2)$ and  $\dot{\A} = 1,2$ is the index of the second $SU(2)$. The upper index is related to the complex conjugate of the lower index by $\rho^1 = \rho_2^*, \rho^2 = -\rho_1^* \, ; \,  \omega^{\dot{1}} = \omega_{\dot{2}}^*, \omega^{\dot{2}} = -\omega_{\dot{1}}^*$.   
These functions in $\mathcal{H}_M$ are assumed to be square integrable on the Euclidean compact support so that $\mathcal{H}_M$ is the Hilbert space.

The element of $\mathcal{H}_M$ is denoted by the same form as Eq.(\ref{eq2.16}) but now $\Psi_+$ and $\Psi_-$ are given by. 
\begin{align}
(\Psi_+)_i &= (\varphi_+, \rho_\A, F_+),
\label{eq3.1} \\
(\Psi_-)_{\bar{i}} &= ( \varphi_-, \omega^{\dot{\A}}, F_- ).
\label{eq3.2}
\end{align} 
These functions are Euclidean counterparts of Eq.(\ref{eq2.1}) and Eq.(\ref{eq2.3}). 
Note that the supersymmetry transformations, Eq.(\ref{eq2.2}) and Eq.(\ref{eq2.4}) are not defined in the Euclidean NCG since the supersymmetry relates scalar functions to spinor functions which are represented by the Lorentz group, $SL(2,C)$ in the Minkowskian signature.  

The element of $\mathcal{A}_+$ which corresponds to Eq.(\ref{eq2.5}) is now given by 
\eq
(u_a)_{ij} = \frac{1}{m_0} \begin{pmatrix}
\varphi_a & 0 & 0 \\
\rho_{a\A} & \varphi_a & 0 \\
F_a & -\rho_a^{*\A} & \varphi_a
\end{pmatrix} \mbox{ } \in \mbox{ } \mathcal{A}_+,
\label{eq3.3}
\eqend
where $\rho_{a\A}$ is $(\frac{1}{2}, 0)$ of $SU(2) \otimes SU(2)$.
Similarly, the element of $\mathcal{A}_-$ corresponding to Eq.(\ref{eq2.8}) is given by
\eq
(\ovl{u}_a)_{\bar{i} \bar{j}} = \frac{1}{m_0} \begin{pmatrix}
\varphi_a^* & 0 & 0 \\
\omega_a^{\dot {\A}} & \varphi_a^* & 0 \\
F_a^* & -\omega^*_{a \dot{\A}} & \varphi_a^*
\end{pmatrix} \mbox{ } \in \mbox{ } \mathcal{A}_-,
\label{eq3.4} 
\eqend
and $\omega_{a}^{\dot{\A}}$ is $(0, \frac{1}{2})$ of $SU(2) \otimes SU(2)$.
The multiplication rule of Eq.(\ref{eq2.6}) also holds by replacing $\psi_{a\A}$ and $\psi^\A_a$ with $\rho_{a\A}$ and $\rho^{*\A}_a$, respectively.
For the algebra $\mathcal{A}_-$,   $\ovl{\psi}_a^{\dot{\A}}$ and $\ovl{\psi}_{a\dot{\A}}$ should be replaced with $\omega^{\dot{\A}}$ and $\omega^*_{\dot{\A}}$, respectively.   These algebras $\mathcal{A}_\pm$ are represented in the Hilbert space $\mathcal{H}_M$ by the relation of the same form as Eq.(\ref{eq2.9}).

In order to obtain the Dirac operator in the Euclidean signature we replace the Minkowskian signature $g^{\mu\nu}$ by $\eta^{\mu\nu}$.
and the Pauli matrices by
\eq
\sigma_E^\mu = ( i\sigma^0,  \sigma^i ) , \mbox{ } \ovl{ \sigma }_E^\mu = ( {i\sigma}^0,  -{\sigma}^i ).
\label{eq3.5}
\eqend
Then from the supersymmetric  Dirac operator (\ref{eq2.11}) and (\ref{eq2.12}) we have  
\begin{align}
\mathcal{D}_M &= -i \begin{pmatrix}
0 & \ovl{\mathcal{D}}_E \\
\mathcal{D}_E & 0
\end{pmatrix},
\label{eq3.6}
\end{align}
where $\mathcal{D}_E$ and $\ovl{\mathcal{D}}_E$ are given by
\begin{align}
\mathcal{D}_E &= \begin{pmatrix}
0 & 0 & 1 \\
0 & i\ovl{\sigma}_E^\mu \partial_\mu & 0 \\
\square_E & 0 & 0 
\end{pmatrix} ,
\label{eq3.7} 
\end{align}
and
\begin{align}
\ovl{\mathcal{D}}_E &= \begin{pmatrix}
0 & 0 & 1 \\
0 & i\sigma^\mu_E \partial_\mu & 0 \\
\square_E & 0 & 0
\end{pmatrix},
\label{eq3.8}
\end{align}
with 
\begin{align}
\square_E &= \eta^{\mu\nu} \partial_\mu \partial_\nu  = \partial_0^2 + \partial_i^2 . 
\label{eq3.9} 
\end{align}
Since $\sigma_E^\mu$ and $\ovl{\sigma}_E^\mu$ are the $(\frac{1}{2}, \frac{1}{2} )$ tensors of $SU(2) \otimes SU(2)$, we have the following relations:
\begin{align}
\rho'_\A &= i (\sigma_E^\mu)_{\A\dot{\A}} \partial_\mu \omega^{\dot{\A}},
\label{eq3.10} \\
\omega'^{\dot{\A}} &= i (\ovl{\sigma}_E^{\mu})^{\dot{\A}\A} \partial_\mu \rho_\A 
\label{eq3.11}.
\end{align}
Therefore, $i\mathcal{D}_M \Psi, \Psi \in \mathcal{H}_M$ is again the element of $\mathcal{H}_M$ and the same relation as Eq.(\ref{eq2.17}) still holds in the Euclidean signature.

Now, the Dirac operator given by Eq.(\ref{eq3.6}) has the following property:
\begin{enumerate}
\item[(a)] $\mathcal{D}_M$ has the real eigenvalues.

Let us compute the eigenvalues of $\mathcal{D}_M$ for the Euclidean $d$-dimensional torus, $M = T^d$ with circumference $2\pi$, where $d$ is the space-time dimension, $d = 4$. After a Fourier transform, 
\begin{align}
\Psi(x) &= \Psi_n \, e^{-i \sum n_i x_i} \mbox{ } (n_i = 0, \pm 1, \cdots),
\label{eq3.12}
\end{align}
the eigenvalue equation reads as follows,
\eq
\mathcal{D}_M(n) \Psi_n = \lambda_n \Psi_n,
\label{eq3.13}
\eqend
where
\begin{align}
\mathcal{D}_M(n) &= \begin{pmatrix}
0 & \ovl{\mathcal{D}}(n) \\
\mathcal{D}(n) & 0
\end{pmatrix}.
\label{eq3.14}
\end{align}
In Eq.(\ref{eq3.14}),  $\mathcal{D}(n)$ and $\ovl{\mathcal{D}}(n)$ are given by
\eq
\mathcal{D}(n) = \begin{pmatrix}
0 & 0 & 0 & -i \\
0 &  n_0 +i n_3 & i n_1+ n_2 & 0 \\
0 & i n_1 - n_2 &  n_0 -i  n_3 & 0 \\
i n^2 & 0 & 0 & 0
\end{pmatrix},
\label{eq3.15}
\eqend
and
\eq
\ovl{\mathcal{D}}(n) = \begin{pmatrix}
0 & 0 & 0 & -i \\
0 &  n_0 -i n_3 & -i n_1- n_2 & 0 \\
0 & -i n_1 + n_2 &  n_0 + i n_3 & 0 \\
i n^2 & 0 & 0 & 0
\end{pmatrix},
\label{eq3.16}
\eqend
where $n^2 = n_0^2 + n_1^2 + n_2^2 + n_3^2$. The characteristic equation amounts to
\begin{align}
\mbox{det}\left| \mathcal{D}_M(n) - \lambda \mathbf{1}_8 \right| &= (\lambda^2 - n^2)^4
\nonumber \\
&= 0,
\label{eq3.17}
\end{align}
which gives the fourthly degenerate eigenvalues of $\mathcal{D}_M$ as
\eq
\lambda_n = \pm \sqrt{n_0^2 + n_1^2 + n_2^2 + n_3^2}.
\label{eq3.18}
\eqend
For large $|\lambda_n|$ there are about $(\pi^2/2) |\lambda_n|^4$ eigenvalues inside the four dimensional ball with the radius $|\lambda_n|$. If we arrange $|\lambda_n|$ in an increasing sequence, we obtain
\eq
|\lambda_n| \approx \left( \frac{2n}{\pi^2} \right)^{1/4},
\label{eq3.19}
\eqend
for large $n$.
\item[(b)] The resolvent $R(\lambda; \mathcal{D}_M) = (\mathcal{D}_M - \lambda\mathbf{1}_8)^{-1}, \lambda \notin \mbox{R}$ of $\mathcal{D}_M$ is compact.

As a matter of fact, for any $\varepsilon > 0$ with sufficiently large $N$ the norm of the resolvent obeys the following relation:

%
%
\begin{align}
||R(\lambda; \mathcal{D}_M)  || &<  | \lambda_{N+1} -\lambda |^{-1} 
\nonumber \\
& < \varepsilon,
\label{eq3.21}.
\end{align}
on the orthogonal  of a $N$ dimensional subspace of $\mathcal{H}_M$.
\item[(c)]  The commutators, $[\mathcal{D}_M, a] = \mathcal{D}_M a -a \mathcal{D}_M $ are bounded for any $a \in \mathcal{A}_M$.

The relations due to the supersymmetry such as Eq.(\ref{eq2.9}) and Eq.(\ref{eq2.15}) in the Minkowskian signature are transfered to the corresponding relations in the Euclidean signature. This implies that if $\Psi$ is an element of $\mathcal{H}_M$, then $a \Psi$ and  $\mathcal{D}_M \Psi$ are also the element of $\mathcal{H}_M$, so that $[\mathcal{D}_M, a]$ is a bounded operator which maps a finite element of the Hilbert space to another finite element.  
\end{enumerate}

A Z/2 grading of the Hilbert space $\mathcal{H}_M$ is given by an operator $\gamma_M$ in $\mathcal{H}_M$ which is defined by
\eq
\gamma_M = \begin{pmatrix}
-1 & 0 \\
0 & 1
\end{pmatrix} ,
\label{eq3.22} 
\eqend
on the basis such that $\gamma_M(\Psi_+)=-1$ and $\gamma_M(\Psi_-)=1$.  Obviously $\gamma_M$ satisfies $\gamma_M^* = \gamma_M, \gamma_M^2 = 1$ and obeys the following relations:
\begin{align}
&\gamma_M a = a \gamma_M, \mbox{  } \forall a \in \mathcal{A}_M,
\label{eq3.23}  \\
&\mathcal{D}_M \gamma_M = - \gamma_M \mathcal{D}_M.
\label{eq3.24} 
\end{align}

In the rest  of this section, let us consider the real structure of the Hilbert space.
First let us define the charge conjugation. 
For the state $\Psi \in \mathcal{H}_M$ the charge conjugate state $\Psi^c$ is given by
\begin{align}
\Psi^c &= \begin{pmatrix}
\Psi_+^c \\
\Psi_-^c  
\end{pmatrix} , 
\label{eq3.25}
\end{align}
and
\begin{align}
(\Psi_+^c)_i &= ( \varphi_+^*, \rho^\A,  F_+^*),
\label{eq3.26} \\
(\Psi_-^c)_{\bar{i}} &= ( \varphi_-^*, \omega_{\dot{\A}},  F_-^* ).
\label{eq3.27}
\end{align} 
Then let us define the antilinear operator $\mathcal{J}_M$ by
\eq
\Psi^c  = \mathcal{J}_M \Psi = C \Psi^*,
\label{eq3.28}
\eqend
so that it is given by
\eq
\mathcal{J}_M = C \otimes * ,
\label{eq3.29}
\eqend
where $C$ is the following charge conjugation matrix;
\eq
C = \left(
\begin{array}{@{\,}ccc|ccc@{\,}}
1 & & & & & \\
 & \varepsilon^{\A\B} & & & \bm{{0}} & \\ 
 & & 1 &  & & \\ \hline
 & & & 1 & & \\
 & \bm{{0}} & &  & \varepsilon_{\dot{\A}\dot{\B}} & \\
  & & &  & & 1
\end{array}
\right),
\label{eq3.30}
\eqend
and $*$ is the complex conjugation (hermitian conjugation for matrices). 
The operator $\mathcal{J}_M$  obeys the following relation:
\begin{align}
\mathcal{J}_M \gamma_M &= \gamma_M \mathcal{J}_M.
\label{eq3.31} 
\end{align}

The real structure $J_M$ is now expressed for the basis of the Hilbert space given by $(\Psi, \Psi^c)^T$ in the following matrix form:
\eq
J_M = \begin{pmatrix}
0 & \mathcal{J}_M^{-1} \\
\mathcal{J}_M & 0
\end{pmatrix}.
\label{eq3.32}
\eqend
On the same basis, the Dirac operator $D_M$ and the Z/2 grading $\Gamma_M$ is expressed by
\eq
D_M = \begin{pmatrix}
\mathcal{D}_M & 0 \\
0 & \mathcal{J}_M \mathcal{D}_M \mathcal{J}_M^{-1}
\end{pmatrix},
\label{eq3.33}
\eqend
and
\eq
\Gamma_M = \begin{pmatrix}
\gamma_M & 0 \\
0 & \gamma_M
\end{pmatrix}.
\label{eq3.34}
\eqend
The real structure defined by Eq.(\ref{eq3.32}) satisfies the following relations:
\begin{align}
J_M^2 &= 1 ,
\label{eq3.35} \\
J_M {D}_M &= {D}_M J_M ,
\label{eq3.36} \\
J_M {\Gamma}_M &= {\Gamma}_M J_M .
\label{eq3.37}
\end{align}

In the Minkowskian signature, we shall define $\mathcal{J}_M$ by the same relation as Eq.(\ref{eq3.28}).  In this case, however, the charge conjugation is defined for Dirac spinors.  A Dirac spinor $\psi$ is composed  of two Weyl spinors, $\chi$ and $\xi$ as
\eq
\psi =  \begin{pmatrix}
\chi_\A \\
\ovl{\xi}^{\dot{\A}}
\end{pmatrix}.
\label{eq3.38}
\eqend
Then, the state $\Psi$ and its charge conjugate state $\Psi^c$ in $\mathcal{H}_M$ is denoted by 
\eq
\Psi = \left(
\varphi_\chi ,
\chi_\A ,
F_\chi ,
\varphi_\xi^* ,
\ovl{\xi}^{\dot{\A}} ,
F_\xi^*
\right)^T,
\label{eq3.39}
\eqend
and
\eq
 \Psi^{c}  = \left( 
\varphi_\xi ,
\xi_\A ,
F_\xi ,
\varphi_\chi^* ,
\ovl{\chi}^{\dot{\A}} ,
F_\chi^*
\right)^T.
\label{eq3.40}
\eqend
The charge conjugation matrix in Eq.(\ref{eq3.29}) is now given by
\eq
{C} = \left(
\begin{array}{@{\,}ccc|ccc@{\,}}
 & & & 1 & 0 & 0   \\
 & {\bf 0} & & 0 & \varepsilon_{\A\B} & 0   \\
 & & & 0 & 0 & 1  \\ \hline
 1 & 0 & 0 & & &  \\
 0 & \varepsilon^{\dot{\A}\dot{\B}} & 0 & & {\bf 0} & \\
 0 & 0 & 1 & & &
\end{array}
\right).
\label{eq3.41}
\eqend

The Z/2 grading  in the Minkowskian signature is defined by
\eq
\gamma_M = \begin{pmatrix}
-i & 0 \\
0 & i
\end{pmatrix},
\label{eq3.42}
\eqend
since the state $\Psi$ and its charge conjugate state $\Psi^c$ in $\mathcal{H}_M$ are given by Eq.(\ref{eq3.39}) and Eq.(\ref{eq3.40}) so that
\begin{align}
\gamma_M(\Psi_+) &= \gamma_M(\Psi_-^*) = \gamma_M(\Psi_+^c) = -i ,
\label{eq3.43} \\
\gamma_M(\Psi_-) &= \gamma_M(\Psi_+^*) = \gamma_M(\Psi_-^c) = i .
\label{eq3.44}
\end{align}
Although $\gamma_M$ and $C$ are anticommuting, $\mathcal{J}_M$ defined by Eq.(\ref{eq3.29}) commutes with $\gamma_M$ and satisfies the same relation as Eq.(\ref{eq3.31}).
As a result, the real structure  $J_M$ as well as  $D_M$ and $\Gamma_M$ in the Minkowskian signature satisfy the same relation as 
 Eq.(\ref{eq3.35})--Eq.(\ref{eq3.37}).
%
%
\section{Internal fluctuations  and vector supermultiplet}
%

In the nonsupersymmetric noncommutative geometry the vector field was introduced as the internal fluctuation on the metric by the following formula,
\begin{align}
A &= \sum a_i [D, b_i]  = i \gamma^\mu \sum a_i \partial_\mu b_i,
\label{eq4.1} \\ 
A_\mu &=  i \sum a_i \partial_\mu b_i,
\label{eq4.2}
\end{align}
here $a_i, b_i \in \mathcal{A}$ and the Dirac operator $D$ is given by $D = i \gamma^\mu \partial_\mu$ \cite{connes7, connes10}. 

In order to introduce  vector fields in the supersymmetric theory we need  two sets of the elements of $\mathcal{A}_+$ and $\mathcal{A}_-$,
\begin{align}
\Pi_+ &= \{ u_a ; a = 1, 2, \cdots, n \} \subset \mathcal{A}_+ ,
\label{eq4.3} \\
{\Pi}_- &=  \{ \ovl{u}_a ; a = 1, 2, \cdots, n \} \subset \mathcal{A}_- ,
\label{eq4.4}
\end{align}
where $u_a$ and $\ovl{u}_a$ are 
given in the matrix form of Eq.(\ref{eq2.5}) and Eq.(\ref{eq2.8}). 
Since the product of chiral (antichiral) supermultiplets is again the chiral (antichiral) supermultiplet, the elements of $\Pi_+ \,({\Pi}_-) $ are chosen such the products of two or more $u_a's \,(\ovl{u}_a's)$ do not belong to $\Pi_+ \, ({\Pi}_-) $ any more.  
Since we consider supersymmetry, we work in the Minkowskian signature so that elements of the algebra obey the supersymmetry transformations given by Eq.(\ref{eq2.2}) and Eq.(\ref{eq2.4}). 

We shall  define the following scalar, spinor and vector fields as the bilinear form of the two component functions in $u_a \in {\Pi}_+$ and $\ovl{u}_a \in {\Pi}_-$;
\begin{align}
&m_0^2 \, C = \sum_a c_a \varphi_a^* \varphi_a ,
\label{eq4.5} \\ 
& m_0^2 \, \chi_\A = -i \sqrt{2} \sum_a c_a \varphi_a^* \psi_{a\A} ,
\label{eq4.6} \\
& m_0^2 \, (M + i N) = -2i \sum_a c_a \varphi_a^* F_a ,
\label{eq4.7} \\
& m_0^2 \, A_\mu = -i \sum_a c_a \left[(\varphi_a^* \partial_\mu \varphi_a - \partial_\mu \varphi_a^* \varphi_a ) \right.
\nonumber \\
& \hspace{4.5cm} \left. - i \ovl{\psi}_{a\dot{\A}} \ovl{\sigma}_\mu^{\dot{\A}\A} \psi_{a\A} \right],
\label{eq4.8} \\
&m_0^2 \,  \lambda_\A = \sqrt{2} i \sum_a c_a \left( F_a^* \psi_{ a\A} - i \sigma^\mu_{\A\dot{\A}} \ovl{\psi}_a^{\dot{\A}} \partial_\mu \varphi_a \right),
\label{eq4.9} \\
&m_0^2 \, D =  
\sum_a c_a \left[ 2  F_a^* F_a-2(\partial^\mu \varphi_a^* \partial_\mu \varphi_a) \right.
\nonumber \\
& \hspace{1cm} \left. + i  \left\{ \partial_\mu \ovl{\psi}_{a\dot{\A}} \ovl{\sigma}^{\mu \dot{\A}\A} \psi_{a\A} -\ovl{\psi}_{a\dot{\A}} \ovl{\sigma}^{\mu \dot{\A}\A} \partial_\mu \psi_{a\A} \right\}  \right] ,
\label{eq4.10}
\end{align}
where $c_a$ are the real coefficients.  Using Eq.(\ref{eq2.2}) and Eq.(\ref{eq2.4}), we can show that these fields have the following transformation property of the vector supermultiplet expressed by
\begin{align}
&\delta_\xi C = i \xi^\A \chi_\A - i \bar{\xi}_{\dot{\A}} \ovl{\chi}^{\dot{\A}} ,
\label{eq4.11} \\
&\delta_\xi \chi_\A = -i \sigma^\mu_{\A\dot{\A}} \bar{\xi}^{\dot{\A}} (-A_\mu + i \partial_\mu C) + \xi_\A (M + iN),
\label{eq4.12} \\
&\frac{1}{2} \delta_\xi (M + i N) = \bar{\xi}_{\dot{\A}} \left( \ovl{\lambda}^{\dot{\A}} + i \bar{\sigma}^{\mu \dot{\A}\A} \partial_\mu \chi_\A \right) ,
\label{eq4.13} \\
&\delta_\xi A^\mu = i \xi^\A \sigma^\mu_{\A\dot{\A}} \ovl{\lambda}^{\dot{\A}} + i \bar{\xi}_{\dot{\A}} \bar{\sigma}^{\mu\dot{\A}\A} \lambda_\A 
\nonumber \\
& \hspace{4cm} + \xi^\A \partial^\mu \chi_\A + \bar{\xi}_{\dot{\A}} \partial^\mu \bar{\chi}^{\dot{\A}},
\label{eq4.14} \\
&\delta_\xi \lambda_\A = {\sigma^{\mu\nu}_\A}^\B \xi_\B (\partial_\mu A_\nu - \partial_\nu A_\mu) + i \xi_\A D ,
\label{eq4.15} \\
&\delta_\xi D = -\xi^\A \sigma^\mu_{\A\dot{\A}} \partial_\mu \ovl{\lambda}^{\dot{\A}} + \bar{\xi}_{\dot{\A}} \bar{\sigma}^{\mu\dot{\A}\A} \partial_\mu \lambda_\A.
\label{eq4.16}
\end{align}
If we express these fields as the superfield, we have
%
\begin{align}
V(x,\theta,\bar{\theta}) &= C + \theta^\A (i\chi_\A) + \bar{\theta}_{\dot{\A}} (-i\chi^{\dot{\A}}) + \theta^\A \sigma^\mu_{\A\dot{\A}}  \bar{\theta}^{\dot{\A}} (-A_\mu)
\nonumber \\
& + \theta\theta \left[ \frac{i}{2} (M+iN) \right] + \bar{\theta}\bar{\theta} \left[ -\frac{i}{2} (M-iN) \right] 
\nonumber \\
& + \theta\theta \bar{\theta}_{\dot{\A}} \left[ i \left( \ovl{\lambda}^{\dot{\A}} + \frac{i}{2} \bar{\sigma}^{\mu\dot{\A}\A} \partial_\mu \chi_\A \right) \right] 
\nonumber \\
&+ \bar{\theta}\bar{\theta} \theta^\A \left[ -i \left( \lambda_\A + \frac{i}{2} \sigma^\mu_{\A\dot{\A}} \partial_\mu \ovl{\chi}^{\dot{\A}} \right) \right] 
\nonumber \\
&+  \theta\theta\bar{\theta}\bar{\theta} \left( \frac{1}{2} D + \frac{1}{4} \square C \right).
\label{eq4.17}
\end{align}

When we define the vector supermultiplet given by Eq.(\ref{eq4.5})--Eq.(\ref{eq4.10}), there is an ambiguity due to the choice of the algebraic elements.  In order to see this, we consider two arbitrary elements of the algebra  given by $u_0 \in \mathcal{A}_+$ and $\ovl{u}_0 \in \mathcal{A}_-$.     It turns out that  the following functions obtained by these elements obey the supersymmetry transformation  of the vector supermultiplet   given by Eq.(\ref{eq4.11})--Eq.(\ref{eq4.16}),
\begin{align}  
& m_0 \, C_0 = \varphi_0 + \varphi_0^*,
\label{eq4.18} \\
& m_0 \, \chi_{0 {\A}} = - i \sqrt{2} \psi_{0 {\A}},
\label{eq4.19} \\
& m_0 \, (M_0+iN_0) = - 2i F_0,
\label{eq4.20} \\
& m_0 \, A_{0 \mu} = - i \partial_\mu ( \varphi_0 - \varphi_0^* ),
\label{eq4.21} \\
& \hspace{6mm} \lambda_{0 \A} = 0,
\label{eq4.22} \\
& \hspace{7mm} D_0 =0.
\label{eq4.23}
\end{align}
Then we can redefine $C, \chi_\A, M, N $ such that
\begin{align}
& C \mbox{ } \rightarrow \mbox{ } C +  C_0 = 0,
\label{eq4.24} \\
& \chi_\A  \mbox{ } \rightarrow \mbox{ } \chi_\A + \chi_{0 {\A}} = 0,
\label{eq4.25} \\
& M + iN \mbox{ } \rightarrow \mbox{ }  (M+M_0) + i (N+N_0) = 0.
\label{eq4.26}
\end{align}
To choose $C, \chi_\A, M $ and  $N $ in the vector supermultiplet to be zero is  called the Wess-Zumino gauge.  This gauge is realized  in Eq.(\ref{eq4.5})--Eq.(\ref{eq4.10}) by the following condition:
\eq
\begin{cases}
\sum_a c_a \varphi^*_a \varphi_a = 0, \\
\sum_a c_a \varphi^*_a \psi_a^\A = 0, \\
\sum_a c_a \varphi^*_a F_a = 0.
\end{cases} 
\label{eq4.27}
\eqend
Hereafter let us call Eq.(\ref{eq4.27}) the Wess-Zumino gauge condition.

Now let us calculate the internal fluctuation of the supersymmetric Dirac operator $\mathcal{D}_M$ given by Eq.(\ref{eq2.11}). The modified Dirac operator is denoted by
\eq
\widetilde{\mathcal{D}}_M = -i \begin{pmatrix}
0 & \widetilde{\ovl{\mathcal{D}}}_{i \bar{j}} \\
\widetilde{{\mathcal{D}}}_{\bar{i} j} & 0
\end{pmatrix}.
\label{eq4.28}
\eqend
  
First, we consider the fluctuation due to $u_a \in \Pi_+ $ and $\bar{u}_a \in {\Pi}_-$.  The contribution to ${\widetilde{\mathcal{D}}}_{\bar{i} j} $ is given by the following matrix form: 
\begin{align}
 V_{\bar{i} j} &= -2 \sum_a c_a (\bar{u}_a)_{\bar{i} \bar{k}} [i\mathcal{D}_M, u_a]_{\bar{k} j}  
\nonumber \\
&=  -2 \sum_a c_a (\bar{u}_a)_{\bar{i} \bar{k}} \, \mathcal{D}_{\bar{k} \ell} \, (u_a)_{\ell j} , 
\label{eq4.29}
\end{align}
and  the contribution to $\widetilde{\ovl{\mathcal{D}}}_{{i} \bar{j}} $ is given by
\begin{align}
\ovl{V}_{i \bar{j}} &=  2 \sum_a c_a (u_a)_{ik} [i\mathcal{D}_M, \bar{u}_a]_{k\bar{j}} 
\nonumber \\
&= 2 \sum c_a (u_a)_{i k} \, \ovl{\mathcal{D}}_{k \bar{\ell}} \, (\bar{u}_a)_{\bar{\ell} \bar{j}}  .
\label{eq4.30}
\end{align}
We shall calculate in the Wess-Zumino gauge. Using the definition of the vector supermultiplet given by Eq.(\ref{eq4.5})--Eq.(\ref{eq4.13}), we obtain the following result:
\begin{align}
 &V_{\bar{i} j} = -  \begin{pmatrix}
0 & 0 & 0 \\
{i}{\sqrt{2}} \ovl{\lambda}^{\dot{\A}} & - \ovl{\sigma}^{\mu\dot{\A}\A} A_\mu & 0 \\
 D + i\partial^\mu A_\mu  + 2i A_\mu \partial^\mu & {i}{\sqrt{2}} \lambda^\A & 0
\end{pmatrix},
\label{eq4.31}
\end{align}
and
\begin{align}
 &\ovl{V}_{i\bar{j} } =
 \begin{pmatrix}
0 & 0 & 0 \\
-{i}{\sqrt{2}} {\lambda}_{{\A}} &  {\sigma}^{\mu}_{\A\dot{\A}} A_\mu & 0 \\
 D - i\partial^\mu A_\mu  - 2i A_\mu \partial^\mu & -{i}{\sqrt{2}} \ovl{\lambda}_{\dot{\A}} & 0
\end{pmatrix}.
\label{eq4.32}
\end{align}

Next, let us consider the additional fluctuation due to $u_{ab} = u_a u_b \in \mathcal{A}_+$ and $\ovl{u}_{ab} = \bar{u}_a \bar{u}_b \in \mathcal{A}_-$, where $a,b = 1, \cdots, n$.  This fluctuation is not contained in the  fluctuation due to $u_a \in \Pi_+ $ and $\ovl{u}_a \in {\Pi}_-$ since  $u_{ab} \notin  \Pi_+$ and  $\ovl{u}_{ab} \notin  {\Pi}_-$. 
The component fields of $u_{ab}$ are expressed by the matrix form of Eq.(\ref{eq2.5}) and  each field is given by
\begin{align}
u_{ab} &=  \{ \varphi_{ab}, \psi_{ab\A}, F_{ab} \} ,
\label{eq4.33} 
\end{align}
where
\begin{align}
\varphi_{ab} &= \frac{1}{m_0} \varphi_a \varphi_b ,
\label{eq4.34} \\
\psi_{ab\A} &= \frac{1}{m_0} (\psi_{a\A} \varphi_b + \varphi_a \psi_{b\A} ),
\label{eq4.35} \\
F_{ab} &= \frac{1}{m_0} ( \varphi_a F_b + F_a \varphi_b - \psi^\A_a \psi_{b\A} ).
\label{eq4.36}
\end{align}
The component fields of $\ovl{u}_{ab}$ are  the complex conjugate functions of  Eq.(\ref{eq4.34})--Eq.(\ref{eq4.36}).  

It turns out that the gauge-covariant form of $\widetilde{\mathcal{D}}_M$ is obtained by considering 
the following fluctuation due to $u_{ab} $ and $\ovl{u}_{ab}$: 
\begin{align}
 V'_{\bar{i} j} &= 2 \sum c_a c_b (\ovl{u}_{ab})_{\bar{i}\bar{k}} [ i\mathcal{D}_M, u_{ab} ]_{k j} 
\nonumber \\
&= 2 \sum c_a c_b (\ovl{u}_{ab})_{\bar{i}\bar{k}} \mathcal{D}_{\bar{k} \ell} (u_{ab})_{\ell j},
\label{eq4.37}
\end{align}
and
\begin{align}
 \ovl{V}'_{{i} \bar{j}}  &= 2 \sum c_a c_b ({u}_{ab})_{{i}{k}} [i \mathcal{D}_M, \ovl{u}_{ab} ]_{k j}
\nonumber \\
&= 2 \sum c_a c_b ({u}_{ab})_{{i}{k}} \ovl{\mathcal{D}}_{{k} \bar{\ell}} (\ovl{u}_{ab})_{ \ovl{\ell} \bar{j}}.
\label{eq4.38} 
\end{align}
Taking into account the Wess-Zumino gauge condition given by
Eq.(\ref{eq4.27}),  we obtain
\eq
V'_{\bar{3} 1} =  \ovl{V}'_{{3} \bar{1}}  =  -A_\mu A^\mu,
\label{eq4.39} 
\eqend
and other matrix elements turn out to be zero.  The fluctuation due to higher order products of $ u_a$ or $\bar{u}_a$ such as $u_{abc} = u_a u_b u_c$ or $\ovl{u}_{abc} = \bar{u}_a \bar{u}_b \bar{u}_c$ vanishes due to the Wess-Zumino gauge condition.  
Thus the total  fluctuation in the Wess-Zumino gauge  amounts to
\begin{align}
V_{\bar{i}j}^{WZ} &= V_{\bar{i}j} + V'_{\bar{i}j} ,
\label{eq4.40} \\
\ovl{V}_{{i}\bar{j}}^{WZ} &= \ovl{V}_{{i}\bar{j}} + \ovl{V}'_{{i}\bar{j}} ,
\label{eq4.41}
\end{align}
and the Dirac operator with fluctuation denoted by Eq.(\ref{eq4.28}) is finally given by
\begin{align}
\widetilde{\mathcal{D}}_{\ovl{i}j} &= {\mathcal{D}}_{\ovl{i}j} +  V^{WZ}_{\bar{i}j}
\nonumber \\
&= 
\begin{pmatrix}
0 & 0 & 1 \\
-i \sqrt{2}  \ovl{\lambda}^{\dot{\A}} & i\ovl{\sigma}^\mu \mathcal{D}_\mu & 0 \\
\mathcal{D}_\mu\mathcal{D}^\mu - D & - i \sqrt{2}  \lambda^\A & 0
\end{pmatrix},
\label{eq4.42}
\end{align}
and
\begin{align}
\widetilde{\mathcal{\ovl{D}}}_{{i}\ovl{j}} &= {\mathcal{\ovl{D}}}_{{i}\ovl{j}} +  \ovl{V}^{WZ}_{{i}\ovl{j}}
\nonumber \\
&= 
\begin{pmatrix}
0 & 0 & 1 \\
-i \sqrt{2}  {\lambda}_{{\A}} & i{\sigma}^\mu \mathcal{D}_\mu & 0 \\
\mathcal{D}_\mu\mathcal{D}^\mu + D & - i \sqrt{2}  \ovl{\lambda}_{\dot{\A}} & 0
\end{pmatrix},
\label{eq4.43}
\end{align}
where $\mathcal{D}_\mu$ is the covariant derivative,
\eq
\mathcal{D}_\mu = \partial_\mu - i A_\mu .
\label{eq4.44}
\eqend
%
%
\section{Internal degrees of freedom and Finite geometry }
%
%
Internal degrees of freedom are introduced in NCG by the finite geometry described by the spectral triple $(\mathcal{A}_F,\mathcal{H}_F,\mathcal{D}_F)$.  Let us consider in this section the color degrees of freedom of quarks and let $\mathcal{A}_F = M_3(\mathbb{C})$, the algebra of  $3 \times 3 $ complex matrices. 

We let $\mathcal{H}_F$ be the Hilbert space with basis of the labels $q_L^a$ and $q_R^a$ of quark-supermultiplets which consist of quarks, squarks and  auxiliary fields.  Here $a$ is the color index, $a = 1,2,3$ and $L, R$ denote the eigenstates of the Z/2-grading $\gamma_F$
, which is defined by
\eq
\gamma_F =  \begin{pmatrix}
-1 & 0 \\
0 & 1
\end{pmatrix}   .
\label{eq5.1}
\eqend
For the basis of $\mathcal{H}_F$ given by
\eq
Q^a = 
\begin{pmatrix}
q^a_L \\
q^a_R
\end{pmatrix} \in \mathcal{H}_F,
\label{eq5.2}
\eqend
we have $\gamma_F(q^a_L) = -1, \text{and } \gamma_F(q^a_R) = 1$.
We shall define the antiquark-supermultiplet states in $\mathcal{H}_F$ as follows;
\eq
\begin{split} 
\left(q^c_a \right)_L &= \left(q^a_R \right)^*,
\\
\left(q^c_a \right)_R &= \left(q^a_L \right)^*,
\end{split}
\label{eq5.3}
\eqend 
and
\eq
Q^c_a = \begin{pmatrix}
\left(q^c_a \right)_L \\
\left(q^c_a \right)_R
\end{pmatrix}  \in \mathcal{H}_F.
\label{eq5.4}
\eqend
Let us define the antilinear operator 
$\mathcal{J}_F$ by
\eq
\mathcal{J}_F =
\begin{pmatrix} 
0 & 1 \\
1 & 0 
\end{pmatrix}  \otimes *,
\label{eq5.5}
\eqend
where $*$ is the complex conjugation.  
Then the antiquark-supermultiplet $Q^c_a$ is related to $Q_a$ by
\eq
Q^c_a = \mathcal{J}_F Q^a.
\label{eq5.6}
\eqend

The Dirac operator ${D}_F$ on the basis $(Q^a,Q^c_a)^T$ is given by
\eq
{D}_F =  \begin{pmatrix}
\mathcal{D}_F & 0 \\
0 & \mathcal{J}_F \mathcal{D}_F \mathcal{J}_F^{-1}   
\end{pmatrix}.
\label{eq5.7}
\eqend
Here $\mathcal{D}_F$ is defined by
\eq
\mathcal{D}_F  = 
\begin{pmatrix}
0 & m^T \\
m & 0
\end{pmatrix},
\label{eq5.8}
\eqend  
and $m$ is the mass matrix with respect to the family index of quark-supermultiplets.

On the same basis, the Z/2-grading $\Gamma_F$ and the real structure  $J_F$ are expressed as follows:
\eq
\Gamma_F = \begin{pmatrix}
\gamma_F & 0 \\
0 & \gamma_F
\end{pmatrix}, 
\label{eq5.9}
\eqend
and
\eq
J_F = \begin{pmatrix}
0 & \mathcal{J}_F^{-1} \\
\mathcal{J}_F & 0
\end{pmatrix}.
\label{eq5.10}
\eqend
Since $\gamma_F$ and $\mathcal{D}_F$ are anticommuting we have
\eq
{D}_F \Gamma_F = - \Gamma_F {D}_F .
\label{eq5.11}
\eqend
Furthermore, $J_F, {D}_F, \text{and } \Gamma_F$ obey the following relations;
\begin{align}
J_F^2 &= 1, 
\label{eq5.12} \\
J_F  {D}_F &= {D}_F J_F ,
\label{eq5.13} \\
J_F \Gamma_F  &= -  \Gamma_F J_F  .
\label{eq5.14}
\end{align}
These relations imply that the K-theoretic dimension of the finite space is 6. This dimension is required to avoid the fermion doubling problem \cite{connes5,barrett}. 

The spectral triple $({A}_0, {H}_0, {D}_0)$ for the product of the space-time manifold $M$ by the finite geometry $F$ is given by
\begin{align}
{A}_0 &= \mathcal{A}_M \otimes  \mathcal{A}_F,
\label{eq5.15} \\
{H}_0  &= \mathcal{H}_M \otimes \mathcal{H}_F,
\label{eq5.16} \\
{D}_0  &= {D}_M  \otimes 1 + \Gamma_M \otimes {D}_F .
\label{eq5.17}
\end{align}
The wave functions of the quark-supermultiplets in $\mathcal{H}$ are 
in $\left(\Psi_+, \Psi_- \right) \otimes \left(q^a_L, q^a_R \right)$.  In order to evade fermion doubling \cite{gracia, lizzi} 
we impose that the physical quark wave functions obey the following condition:
\eq
\gamma = \gamma_M \gamma_F  = 1.
\label{eq5.18}
\eqend
Then for the left-handed quark-supermultiplet we have
\begin{align} 
\Psi^a_L(x)  &= q^a_L \otimes \left(\Psi_{+}(x)\right)_i 
\nonumber \\
&= q^a_L \otimes \left(\varphi_+(x), \psi_{+\A}(x), F_+(x) \right)  ,
\label{eq5.19}
\end{align}
in the Minkowskian signature and the wave functions of quark, squark, and auxiliary field amount to
\begin{align}
q^a_{L\A} (x) &= q^a_L \otimes \psi_{+\A} (x),
\label{eq5.20} \\
\tilde{q}^a_L(x) &= q^a_L \otimes \varphi_+ (x) ,
\label{eq5.21} \\
F^a_L(x) &= q^a_L \otimes F_+ (x).
\label{eq5.22}
\end{align}
For the right-handed quark-supermultiplet we have
\eq
\Psi^a_R(x)  = q^a_R \otimes \left(\Psi_-(x)\right)_{\bar{i}} ,
\label{eq5.23}
\eqend
and the wave functions of the component fields are given by
\begin{align}
q_{R}^{a\dot{\A}} (x) &= q^a_R \otimes \psi_{-}^{\dot{\A}} (x),
\label{eq5.24} \\
\tilde{q}^a_R(x) &= q^a_R \otimes \varphi_- (x)  ,
\label{eq5.25} \\
F^a_R(x) &= q^a_R \otimes F_- (x).
\label{eq5.26}
\end{align}
  
The elements of the algebra ${A}_0$ given by Eq.(\ref{eq5.15}) are the matrix-valued supermultiplet, i.e., the component functions of $u_a$ and $ \ovl{u}_a$ 
in Eq.(\ref{eq4.3}) and Eq.(\ref{eq4.4}), respectively, are  $3 \times 3 $ complex matrix  functions. Then the  functions in the vector supermultiplet defined by Eq.(\ref{eq4.5})--Eq.(\ref{eq4.10}) are also   $3 \times 3 $ complex matrix  functions.  In the Wess-Zumino gauge we consider $A_\mu, \lambda_\A$ and $D$.  Among them,  $A_\mu$ and  $D$ are hermitian so that they are parametrized by
\begin{align}
A_\mu(x) &= \sum_{\ell=0,\cdots,8} A_\mu^\ell (x) \,
\frac{t_\ell}{2},
\label{eq5.27} \\
D(x) &= \sum_{\ell=0,\cdots,8} D^\ell(x) \,
\frac{t_\ell}{2},
\label{eq5.28} 
\end{align}
where  $A_\mu^\ell(x)$ and  $D^\ell(x)$ are real functions and $t_\ell$ are  $3 \times 3 $ Gell-Mann matrices.
The complex matrix function $\lambda_\A$ is expressed  by
\eq
\lambda_\A(x) = \sum_{\ell=0,\cdots,8} \lambda_\A^\ell (x) \,
\frac{t_\ell}{2},
\label{eq5.29} 
\eqend
and $\lambda_\A^\ell (x)$ are  complex functions.

Since the vector supermultiplet originates from the fluctuation of the metric, the matrix-valued functions $A_\mu, \lambda_\A$ and  $D$ should be traceless because the trace part does not affect the metric \cite{connes7}.  This amounts to remove $t_0$ from the summation in Eq.(\ref{eq5.27})--Eq.(\ref{eq5.29}).   As a result these functions have the same form as the adjoint representation of $SU(3)$. In the next section we shall show that the vector supermultiplet actually satisfies the super Yang-Mills action with the $SU(3)$ gauge symmetry.  
%
%
\section{supersymmetric QCD and the spectral action principle}
%
%
In the Euclidean NCG models without supersymmetry, the bosonic part of the action is obtained by the spectral action principle, which asserts that the action depends only on the spectrum of the squared Dirac operator.  
In our noncommutative geometric approach to supersymmetry, we have
derived the supersymmetric Dirac operator $\widetilde{\mathcal{D}}_M$ given by Eq.(\ref{eq4.28}) with Eq.(\ref{eq4.42}) and Eq.(\ref{eq4.43}). 

Let us show that the supersymmetric action for the vector supermultiplet will be obtained by the spectral action in the heat kernel expansion of the elliptic operator $P$ :
\eq
\textrm{Tr}_{L^2}  \, f \left( P \right)
\simeq \sum_{n \geq 0} c_{n} \, a_{n} \left( P \right),
\label{eq6.1}
\eqend 
where $f(x)$ is an auxiliary  smooth function on a smooth compact Riemannian manifold $M$ without boundary of dimension 4 \cite{gilkey}.
Since the contribution to $P$ from the antiparticles is the same as that of the particles, we consider only the contribution from the particles. Then 
the elliptic operator $P$ in our case is given by the square of the Wick rotated Euclidean Dirac operator $\widetilde{\mathcal{D}}_0$,  
\eq
\widetilde{\mathcal{D}}_0 =  \widetilde{\mathcal{D}}_M  +  \gamma_M \otimes \mathcal{D}_F ,
\label{eq6.2}
\eqend
where $\widetilde{\mathcal{D}}_M$ is obtained from Eq.(\ref{eq4.28}), Eq.(\ref{eq4.42}) and Eq.(\ref{eq4.43}) by a replacement of $g^{\mu\nu} \rightarrow \eta^{\mu\nu}, \sigma^\mu \rightarrow \sigma_E^\mu, \ovl{\sigma}^\mu \rightarrow \ovl{\sigma}^\mu_E$. 
Note that the internal fluctuation on $\mathcal{D}_F $ is absent in the QCD model.

The elliptic operator $P$ is expanded into the following form:
\begin{align}
P &= 
 - \left( \eta^{\mu\nu} \partial_\mu \partial_\nu + \mathbb{A}^\mu \partial_\mu + \mathbb{B} \right).
 \label{eq6.3}
\end{align}
The heat kernel coefficients $a_n$ in Eq.(\ref{eq6.1}) are found in \cite{gilkey}.  
They vanish for $n$ odd,  and the first three $a_n$'s for $n$ even in the flat space are given by
\begin{align}
a_{0} \left( P \right) &= \frac{1}{16\pi^2} \int_M dx^4  \, \textrm{tr}_V (\mathbb{I}),
\label{eq6.4} \\
a_{2} \left( P \right) &= \frac{1}{16\pi^2} \int_M dx^4  \, \textrm{tr}_V (\mathbb{E}),
\label{eq6.5} \\
a_{4} \left( P \right) &= \frac{1}{32\pi^2} \int_M dx^4  \, 
\nonumber \\
&\hspace{1cm} \times \textrm{tr}_V \left( \mathbb{E}^2 + \frac{1}{3} {\mathbb{E}_{;\mu}}^\mu + \frac{1}{6} \Omega_{\mu\nu} \Omega^{\mu\nu} \right),
\label{eq6.6}
\end{align}
where $\mathbb{E}$ and the bundle curvature $\Omega^{\mu\nu}$ in the flat space are defined as follows;
\begin{align}
\mathbb{E} &= \mathbb{B} - \left( \partial_\mu \omega^\mu + \omega_\mu \omega^\mu \right),
\label{eq6.7} \\
\Omega^{\mu\nu} &= \partial^\mu \omega^\nu - \partial^\nu \omega^\mu + [\omega^\mu, \omega^\nu ],
\label{eq6.8}
\\
\omega^\mu &= \frac{1}{2}  \mathbb{A}^\mu .
\label{eq6.9}
\end{align}

The coefficients $c_{n} $ in Eq.(\ref{eq6.1}) depend on the functional form of $f(x) $.  If $f(x) $ is flat near 0, it turns out that $c_{2k} = 0 $ for $k \geq  3$ and the heat kernel expansion (\ref{eq6.1}) terminates at $n = 4$ \cite{connes8}.

In Eq.(\ref{eq6.4})--Eq.(\ref{eq6.6}), $\textrm{tr}_V$ denotes the trace over the vector bundle $V$. As for the supersymmetric theory we consider here, sections of the vector bundle $V$ are smooth functions bearing indices which correspond to internal and spin degrees of freedom of the chiral or antichiral supermultiplets. 
For the spin degrees of freedom, $\textrm{tr}_V$ is the supertrace defined by 
\begin{align}
\textrm{Str} \, O &= \sum_i \langle i | (-1)^{2s} O | i \rangle
\nonumber \\
&= \sum_b \langle b | O | b \rangle - \sum_f \langle f | O | f \rangle,
\label{eq6.10}
\end{align}
where $s$ is the spin angular momentum and the states $| b \rangle$ and $| f \rangle$ stand for bosonic and fermionic states, respectively.

Now let us calculate the spectral action for $\widetilde{\mathcal{D}}_0^2$ where $\widetilde{\mathcal{D}}_0$ is given by Eq.(\ref{eq6.2}). 
 In the contribution to the spectral action from $\widetilde{\mathcal{D}}_0^2$,  
the terms including $\mathcal{D}_F $ vanish since ${\widetilde{\mathcal{D}}_M}$ anticommutes with $\gamma_M$ and 
\begin{align}
\textrm{Str} \, \gamma_M^2 &= \textrm{Str} \, 1 = 0.
\label{eq6.11}
\end{align}
Thus, we consider the following elliptic operator $P$ in the Euclidean signature,
\eq
P = {\widetilde{\mathcal{D}}_M^2}{} = \begin{pmatrix}
P_+ & 0 \\
0 & P_-
\end{pmatrix}.
\label{eq6.12}
\eqend
By making use of the Wick-rotated expression of Eq.(\ref{eq4.42}) and Eq.(\ref{eq4.43}), $P_\pm$ amounts to
\begin{widetext}
\begin{align}
P_+ &= - \tilde{\ovl{\mathcal{D}}}_E \tilde{\mathcal{D}}_E
\nonumber \\
 &= - \begin{pmatrix}
\mathcal{D}_\mu  \mathcal{D}^\mu - D & - i\sqrt{2} \lambda^\B & 0 \\
\sqrt{2} \sigma^\mu_{E \A\dot{\A}} (\mathcal{D}_\mu \ovl{\lambda}^{\dot{\A}})  + \sqrt{2} \sigma^\mu_{E \A\dot{\A}} \ovl{\lambda}^{\dot{\A}} \mathcal{D}_\mu & \mathcal{D}_\mu  \mathcal{D}^\mu \delta_\A^\B + i \sigma^{\mu\nu\B}_{E \A} F_{\mu\nu} & - i\sqrt{2} \lambda_\A \\
-2 \ovl{\lambda}_{\dot{\A}} \ovl{\lambda}^{\dot{\A}} &  \sqrt{2} \ovl{\lambda}_{\dot{\A}} \ovl{\sigma}_E^{\mu\dot{\A}\B} \mathcal{D}_\mu & \mathcal{D}_\mu  \mathcal{D}^\mu + D
\end{pmatrix},
\label{eq6.13}
\end{align}
and
\begin{align}
P_- &= - \tilde{\mathcal{D}}_E \tilde{\ovl{\mathcal{D}}}_E 
\nonumber \\
&=  -\begin{pmatrix}
\mathcal{D}_\mu  \mathcal{D}^\mu + D & - \sqrt{2} i \ovl{\lambda}_{\dot{\B}} & 0 \\
\sqrt{2} \ovl{\sigma}_E^{\mu\dot{\A}\A} (\mathcal{D}_\mu \lambda_\A) + \sqrt{2} \ovl{\sigma}_E^{\mu\dot{\A}\A} \lambda_\A \mathcal{D}_\mu  & \mathcal{D}_\mu  \mathcal{D}^\mu \delta^{\dot{\A}}_{\dot{\B}} + i {\ovl{\sigma}_E^{\mu\nu\dot{\A}}}_{\dot{\B}} F_{\mu\nu} & - \sqrt{2} i \ovl{\lambda}^{\dot{\A}} \\
- 2 \lambda^\A \lambda_\A & \sqrt{ 2} \lambda^\A \sigma^\mu_{E \A\dot{\B}} \mathcal{D}_\mu & \mathcal{D}_\mu  \mathcal{D}^\mu - D
\end{pmatrix}.
\label{eq6.14}
\end{align}
\end{widetext}
In Eq.(\ref{eq6.13}) and Eq.(\ref{eq6.14}), $\sigma^{\mu\nu}_{E } $ and $\ovl{\sigma}^{\mu\nu}_{E } $ are defined by
\begin{align}
\sigma^{\mu\nu}_{E } &= \left( i \sigma^{0j}, \sigma^{ij} \right),
\label{eq6.15} \\
\ovl{\sigma}^{\mu\nu}_{E } &= \left( i \ovl{\sigma}^{0j}, \ovl{\sigma}^{ij} \right),
\label{eq6.16}
\end{align}
and
\begin{align}
{\sigma_\A^{\mu\nu}}^\B &= \frac{1}{4} \left(  \sigma^{\mu}_{\A\dot{\A}} \ovl{\sigma}^{\nu\dot{\A}\B} - \sigma^{\nu}_{\A\dot{\A}} \ovl{\sigma}^{\mu\dot{\A}\B} \right),
\label{eq6.17} \\
{\ovl{\sigma}^{\mu\nu\dot{\A}}}_{\dot{\B}} &= \frac{1}{4} \left(  \ovl{\sigma}^{\mu\dot{\A}\B}\sigma^{\nu}_{\B\dot{\B}} -  \ovl{\sigma}^{\nu\dot{\A}\B} \sigma^{\mu}_{\B\dot{\B}}  \right).
\label{eq6.18}
\end{align}

The fields $A_\mu, \lambda_\A, D $ 
are the $3 \times 3$ matrices as shown by Eq.(\ref{eq5.27})--Eq.(\ref{eq5.29}). They turn out to be the gauge, gaugino and auxiliary fields.
The covariant derivative on spinors, say, ${\lambda}_{\A}$ is given by
\eq
\mathcal{D}_\mu {\lambda}_{{\A}} =  \partial_\mu {\lambda}_{{\A}} - i[A_\mu, {\lambda}_{{\A}}] ,
\label{eq6.19}
\eqend
and  $F_{\mu\nu} $ is the field strength defined by
\begin{align}
F_{\mu\nu} &= i  [\mathcal{D}_\mu, \mathcal{D}_\nu] 
\nonumber \\
&= \partial_\mu A_\nu - \partial_\nu A_\mu - i [ A_\mu, A_\nu ].
\label{eq6.20}
\end{align}

We expand $P_\pm$ in the form given by Eq.(\ref{eq6.3}).
Using the formulae given by Eq.(\ref{eq6.6}) and Eq.(\ref{eq6.8}), we obtain the following expressions;
\begin{widetext}
\begin{align}
\mathbb{E}_+ &= \mathbb{B}_+ - \left( \partial_\mu \omega_+^\mu + \omega_{+\mu} \omega_+^\mu \right)
= \begin{pmatrix}
-D & -i \sqrt{2} \lambda^\B & 0 \\
\frac{1}{\sqrt{2}} \sigma^\mu_{E \A\dot{\A}} (\mathcal{D}_\mu \ovl{\lambda}^{\dot{\A}}) &  i \sigma^{\mu\nu\B}_{E \A} F_{\mu\nu} & -i \sqrt{2} \lambda_\A \\
-2 \ovl{\lambda}_{\dot{\A}} \ovl{\lambda}^{\dot{\A}} & -\frac{1}{\sqrt{2}} (\mathcal{D}_\mu \ovl{\lambda}_{\dot{\A}}) \ovl{\sigma}^{\mu\dot{\A}\B}_E & D
\end{pmatrix},
\label{eq6.21} 
\end{align}
and
\begin{align}
\mathbb{E}_- &= \mathbb{B}_- - \left( \partial_\mu \omega_-^\mu + \omega_{-\mu} \omega_-^\mu \right)
= \begin{pmatrix}
D & -i \sqrt{2} \,\ovl{\lambda}_{\dot{\B}} & 0 \\
\frac{1}{\sqrt{2}} \ovl{\sigma}_E^{\mu\dot{\A}\A} (\mathcal{D}_\mu {\lambda}_{\A}) &  i {\ovl{\sigma}_E^{\mu\nu\dot{\A}}}_{\dot{\B}} F_{\mu\nu} & -i \sqrt{2} \,\ovl{\lambda}^{\dot{\A}} \\
- 2 \lambda^\A \lambda_\A & -\frac{1}{\sqrt{2}} (\mathcal{D}_\mu {\lambda}^{{\A}}) {\sigma}^{\mu}_{E\A\dot{\B}} & - D
\end{pmatrix} .
\label{eq6.22}
\end{align}
The bundle curvature $\Omega_\pm^{\mu\nu}$ given by Eq.(\ref{eq6.7}) amounts to
\begin{align}
\Omega_+^{\mu\nu} &= \begin{pmatrix}
-i F^{\mu\nu} & 0 & 0 \\
\frac{1}{\sqrt{2}} [\sigma^\nu_{E\A\dot{\A}} (\mathcal{D}^\mu \ovl{\lambda}^{\dot{\A}}) - \sigma^\mu_{E\A\dot{\A}} (\mathcal{D}^\nu \ovl{\lambda}^{\dot{\A}})] & -i F^{\mu\nu} \delta_\A^\B & 0 \\
0 & \frac{1}{\sqrt{2}} [(\mathcal{D}^\mu \ovl{\lambda}_{\dot{\A}}) \ovl{\sigma}^{\nu\dot{\A}\B}_E - (\mathcal{D}^\nu \ovl{\lambda}_{\dot{\A}}) \ovl{\sigma}_E^{\mu\dot{\A}\B}] & -i F^{\mu\nu}
\end{pmatrix},
\label{eq6.23} \\
\Omega_-^{\mu\nu} &= \begin{pmatrix}
-i F^{\mu\nu} & 0 & 0 \\
\frac{1}{\sqrt{2}} [\ovl{\sigma}_E^{\nu\dot{\A}{\A}} (\mathcal{D}^\mu {\lambda}_{{\A}}) - \ovl{\sigma}_E^{\mu\dot{\A}{\A}} (\mathcal{D}^\nu {\lambda}_{{\A}})] & -i F^{\mu\nu} \delta_{\dot{\A}}^{\dot{\B}} & 0 \\
0 & \frac{1}{\sqrt{2}} [(\mathcal{D}^\mu {\lambda}^{{\A}}) {\sigma}^{\nu}_{E{\A}\dot{\B}} - (\mathcal{D}^\nu {\lambda}^{{\A}}) {\sigma}^{\mu}_{E{\A}\dot{\B}}] & -i F^{\mu\nu}
\end{pmatrix}.
\label{eq6.24}
\end{align}

From Eq.(\ref{eq6.21}) we have
\begin{align}
\textrm{Str}  \, \mathbb{E}_+^2 &= \textrm{Tr} \left[ D^2 -i \lambda^\B \sigma^\mu_{E\B \dot{\B}} (\mathcal{D}_\mu \ovl{\lambda}^{\dot{\B}}) \right] 
\nonumber \\
& - \textrm{Tr} \left[ -i \sigma^\mu_{E\A\dot{\A}} (\mathcal{D}_\mu \ovl{\lambda}^{\dot{\A}} ) \lambda^\A - \sigma^{\mu\nu \B}_{E\A} \sigma^{\lambda\kappa \A}_{E\B} F_{\mu\nu} F_{\lambda\kappa} + i \lambda_\A (\mathcal{D}_\mu \ovl{\lambda}_{\dot{\B}} ) \ovl{\sigma}_E^{\mu\dot{\B}\A} \right]
\nonumber \\
& + \textrm{Tr} \left[ i (\mathcal{D}_\mu \ovl{\lambda}_{\dot{\A}} ) \ovl{\sigma}^{\mu\dot{\A}\B}_E \lambda_\B + D^2 \right] 
\nonumber \\
&= \textrm{Tr} \left[ 2 D^2 - 4i \ovl{\lambda}_{\dot{\B}}\ovl{\sigma}_E^{\mu\dot{\B}\B} (\mathcal{D}_\mu\lambda_\B) - F_{\mu\nu} F^{\mu\nu} - \frac{i}{2} \varepsilon^{\mu\nu\lambda\kappa} F_{\mu\nu} F_{\lambda\kappa} \right].
\label{eq6.25}
\end{align}
Here the trace is taken over the $3 \times 3$ matrices of the color degrees of freedom. 
As for the antichiral sector given by Eq.(\ref{eq6.22}) we have the following result:
\eq
\textrm{Str}  \, \mathbb{E}_-^2 =  \textrm{Tr} \left[ 2 D^2 - 4i \ovl{\lambda}_{\dot{\B}}\ovl{\sigma}_E^{\mu\dot{\B}\B} (\mathcal{D}_\mu\lambda_\B) - F_{\mu\nu} F^{\mu\nu} + \frac{i}{2} \varepsilon^{\mu\nu\lambda\kappa} F_{\mu\nu} F_{\lambda\kappa} \right] .
\label{eq6.26}
\eqend
Equation (\ref{eq6.25}) and Eq.(\ref{eq6.26}) give the following expression:
\begin{align}
\textrm{tr}_V  (\mathbb{E}^2) &= \textrm{Str}  \, \mathbb{E}_+^2 + \textrm{Str}  \, \mathbb{E}_-^2 
\nonumber \\
&= 2 \textrm{Tr} \left[ 2 D^2 - 4i \ovl{\lambda}_{\dot{\B}}\ovl{\sigma}_E^{\mu\dot{\B}\B} (\mathcal{D}_\mu\lambda_\B) - F_{\mu\nu} F^{\mu\nu} \right] .
\label{eq6.27}
\end{align}
The supertrace of $\Omega_{\pm\mu\nu}\Omega_\pm^{\mu\nu}$ amounts to
\begin{align}
\textrm{Str}  \, {\Omega}_{\pm\mu\nu} {\Omega}_\pm^{\mu\nu} &= \textrm{Tr} [-F_{\mu\nu} F^{\mu\nu} ] -  \textrm{Tr} [-F_{\mu\nu} F^{\mu\nu} {\bf 1}_2] + \textrm{Tr} [-F_{\mu\nu} F^{\mu\nu} ]
\nonumber \\
&= 0.
\label{eq6.28}
\end{align}
\end{widetext}

Now that we are ready to calculate the heat kernel coefficients.  From Eq.(\ref{eq6.3}) we obtain
\eq
a_0 = 0,
\label{eq6.29}
\eqend
since the number of freedom of the bosonic sector is equal to the number of freedom of the fermionic sector due to the supersymmetry, so that 
$\textrm{Str} \,\mathbb{I} = 0$.  Equation (\ref{eq6.29}) indicates that the cosmological constant in the supersymmetric  
theory vanishes. The coefficient $a_2$ also vanishes,
\eq
a_2 = 0,
\label{eq6.30}
\eqend
since $D$ and $F_{\mu\nu}$ in $\textrm{tr}_V(\mathbb{E})$ are traceless $3 \times 3$ matrices with respect to the color degrees of freedom.
Finally, Eq.(\ref{eq6.27}) and Eq.(\ref{eq6.28}) give
\begin{align}
a_{4}  &= \frac{1}{16 \pi^2} \int_M dx^4  
\nonumber \\
& \times \textrm{Tr} \left[ 2 D^2 - 4i \ovl{\lambda}_{\dot{\B}}\ovl{\sigma}_E^{\mu\dot{\B}\B} (\mathcal{D}_\mu\lambda_\B) - F_{\mu\nu} F^{\mu\nu} \right] ,
\label{eq6.31}
\end{align}
since $\textrm{tr}_V ({\mathbb{E}_{;\mu}}^\mu) = 0$.  

The Euclidean super Yang-Mills action $I_E$ is now given by
\eq
I_E = \textrm{Tr}_{L^2}  \, f \left( \widetilde{\mathcal{D}}_M^2 \right) = f_4 \,a_4.
\label{eq6.32}
\eqend
In order to obtain the physical action we change the signature back to the Minkowskian, $\eta^{\mu\nu} \rightarrow g^{\mu\nu}$ and rescale the vector supermultiplet as $\{ A_\mu, \lambda_\A, D \} \rightarrow  \{ g_c A_\mu, g_c \lambda_\A, g_c D \}$, where $g_c$ turns out to be the gauge coupling constant. After this procedure we have the following super Yang-Mills action:
\begin{align}
& I_{SYM}  = \int_M dx^4  \, 
\nonumber \\
& \hspace{4mm} \times \textrm{Tr} \left[  - \frac{1}{2} F_{\mu\nu} F^{\mu\nu} - 2i \ovl{\lambda}_{\dot{\B}}\ovl{\sigma}^{\mu\dot{\B}\B} (\mathcal{D}_\mu\lambda_\B) + D^2 \right] ,
\label{eq6.33}
\end{align} 
where we fixed the constant $f_4$ such that
\eq
\frac{f_4}{8\pi^2} = \frac{1}{g_c^2},
\label{eq6.34}
\eqend
and normalized the matrices $t_\ell$ in Eq.(\ref{eq5.27})--Eq.(\ref{eq5.29}) as
\eq
\textrm{Tr} \left( t_\ell \, t_m \right) = 2 \, \delta_{\ell m}.
\label{eq6.35}
\eqend

The final step is to derive the action integral including quarks.  Here we work in the Minkowskian signature. 
Let us first define the supersymmetric product of two chiral  supermultiplets $\Psi_+$ and $\Psi_-^\dagger$ by 
\begin{align}
(\Psi_-, \Psi_+)  
&= \int_M dx^4  \, \Psi^\dagger_- \Gamma_0 \Psi_+,
\label{eq6.36}
\end{align}
where $\Gamma_0$ is the constant matrix such that
\begin{align}
\Psi^\dagger_- \Gamma_0 \Psi_+ &= \left(\varphi_-^*, (\psi_-^*)^\A, F_-^* \right) \begin{pmatrix}
0 & 0 & 1 \\
0 & -1 & 0 \\
1 & 0 & 0
\end{pmatrix}
\begin{pmatrix}
\varphi_+ \\
\psi_{+\A} \\
F_+
\end{pmatrix}
\nonumber \\
&= \varphi_-^* F_+ + F_-^* \varphi_+ - (\psi_-^*)^\A \psi_{+\A}.
\label{eq6.37}
\end{align}
This expression is the F-term of the product of two chiral  supermultiplets and it transforms into a space derivative under the supersymmetry transformation given by Eq.(\ref{eq2.2}).  The supersymmetric product of two antichiral  supermultiplets is also defined in the same way. 

The kinetic parts of the actions for the chiral supermultiplet $\Psi_+ = (\varphi_+, \psi_{+\A}, F_+)$ and antichiral  supermultiplet $\Psi_- = (\varphi_-, \psi_-^{\dot{\A}}, F_-)$ are obtained by the use of the following formulae:
\begin{widetext}
\begin{align}
\Psi_+^\dagger \Gamma_0 \widetilde{\mathcal{D}} \Psi_+
&= -\mathcal{D}_\mu \varphi_+^* \mathcal{D}^\mu \varphi_+  - i {\psi}^*_{+\dot{\A}} \ovl{\sigma}^{\mu\dot{\A}\A}  \mathcal{D}_\mu \psi_{+\A} - \sqrt{2} i \left(\varphi_+^* \lambda^\A \psi_{+\A} - {\psi}^*_{+\dot{\A}} \ovl{\lambda}^{\dot{\A}}  \varphi_+ \right) - \varphi_+^* D \varphi_+ + F_+^* F_+,
\label{eq6.38}
\\
\Psi_-^\dagger \Gamma_0 \widetilde{\ovl{\mathcal{D}}} \Psi_-
&= -\mathcal{D}_\mu \varphi_-^* \mathcal{D}^\mu \varphi_-  - i {\psi}_-^{*{\A}} {\sigma}^{\mu}_{\A\dot{\A}}  \mathcal{D}_\mu \psi_-^{\dot{\A}} - \sqrt{2} i \left(\varphi_-^* \ovl{\lambda}_{\dot{\A}} \psi_-^{\dot{\A}} - {\psi}_-^{{*\A}} {\lambda}_{{\A}}  \varphi_- \right) + \varphi_-^* D \varphi_- + F_-^* F_-,
\label{eq6.39}
\end{align}
where $\widetilde{\mathcal{D}}$ and $\widetilde{\ovl{\mathcal{D}}}$ are given by Eq.(\ref{eq4.42}) and Eq.(\ref{eq4.43}).
\end{widetext}

The wave function of quarks, $\Psi$ consists of the chiral- and antichiral-supermultiplets defined by Eq.(\ref{eq5.20})--Eq.(\ref{eq5.22}) and Eq.(\ref{eq5.24})--Eq.(\ref{eq5.26}), respectively. They are denoted by
\begin{align}
\Psi_L^a &= \left( \tilde{q}_L^a, q_{L\A}^a , F_L^a \right),
\label{eq6.40} \\
\Psi_R^a &= \left( \tilde{q}_R^a, q_{R}^{a \dot{\A} } , F_R^a \right),
\label{eq6.41} 
\end{align}
where $a$ is the color index.  The spectral action principle asserts that the action for quarks is obtained by the total Dirac operator $\widetilde{\mathcal{D}}_0$ so that
\begin{align}
I_{\textrm{quark}} &= \left( \Psi, i \widetilde{\mathcal{D}}_0 \Psi \right)
\nonumber \\
&= \left( \Psi, i \widetilde{\mathcal{D}}_M \Psi \right) + \left( \Psi,  \gamma_M \otimes \mathcal{D}_F \Psi \right).
\label{eq6.42}
\end{align}
Here $\gamma_M$ is given by Eq.(\ref{eq3.42}).

The first term of Eq.(\ref{eq6.42}) is the kinetic part of the action, which is given by
\begin{align}
I_{\textrm{kinetic}} &= \left( \Psi, i \widetilde{\mathcal{D}}_M \Psi \right)
\nonumber \\
&= \left( \Psi_L, \widetilde{\mathcal{D}} \Psi_L \right) + \left( \Psi_R, \widetilde{\ovl{\mathcal{D}}} \Psi_R \right).
\label{eq6.43}
\end{align}
The second term of Eq.(\ref{eq6.42}) turns out to be the mass term of quarks: 
\eq
I_{\textrm{mass}} = \left( \Psi,  \gamma_M \otimes \mathcal{D}_F \Psi \right).
\label{eq6.44}
\eqend

As for the kinetic part of the action, using Eq.(\ref{eq6.38}) and Eq.(\ref{eq6.39}) we can confirm that
\begin{widetext}
\begin{align}
I_L &= \left( \Psi_L, \widetilde{\mathcal{D}} \Psi_L \right) 
\nonumber \\
&= \int_M dx^4  \, \left[  -\mathcal{D}_\mu \tilde{q}_L^* \mathcal{D}^\mu \tilde{q}_L - i \ovl{q}_{L\dot{\A}} \ovl{\sigma}^{\mu\dot{\A}\A}  \mathcal{D}_\mu q_{L\A} - \sqrt{2} i g_c \left(\tilde{q}_L^* \lambda^\A q_{L\A} - \ovl{q}_{L\dot{\A}} \ovl{\lambda}^{\dot{\A}}  \tilde{q}_L \right) - \tilde{q}_L^* D \tilde{q}_L + F_L^* F_L  
 \right],
\label{eq6.45} 
\end{align}
for the left-handed quark supermultiplet and
\begin{align}
I_R &= \left( \Psi_R, \widetilde{\ovl{\mathcal{D}}} \Psi_R \right) 
\nonumber \\
&= \int_M dx^4  \, \left[  -\mathcal{D}_\mu \tilde{q}_R^* \mathcal{D}^\mu \tilde{q}_R - i \ovl{q}_{R}^{{\A}} {\sigma}^{\mu}_{\A\dot{\A}}  \mathcal{D}_\mu q_{R}^{\dot{\A}} - \sqrt{2} i g_c \left(\tilde{q}_R^* \ovl{\lambda}_{\dot{\A}} q_{R}^{\dot{\A}} - \ovl{q}_{R}^{{\A}} {\lambda}_{{\A}}  \tilde{q}_R \right) + \tilde{q}_R^* D \tilde{q}_R + F_R^* F_R
 \right],
\label{eq6.46}
\end{align}
for the right-handed quark supermultiplet.
\end{widetext}

Finally let us give the mass term of quarks. From Eq.(\ref{eq6.44}) this part of the action amounts to
\begin{align}
I_{\textrm{mass}} &= 
 \left( \Psi_L, i m^T \Psi_R \right) - \left( \Psi_R, i m \Psi_L \right),
\label{eq6.47} 
\end{align}
where we recall that $\mathcal{D}_F$ is given by Eq.(\ref{eq5.8}). We redefine here the phase of $\Psi_L$ as $\Psi_L \rightarrow i \Psi_L$, then we have  
\begin{align}
I_{\textrm{mass}} &= \left( \Psi_R,  m \Psi_L \right) + \textrm{h.c.}
\nonumber \\
&=  \int_M dx^4  \, 
\nonumber \\
& \hspace{0mm} \times \left[ \tilde{q}_R^* \,m\,  F_L  + F_R^* \,m\, \tilde{q}_L - \ovl{q}_{R}^{{\A}} \,m\, q_{L\A} + \textrm{h.c.} \right] .
\label{eq6.48}
\end{align}

To summarize we have shown that the supersymmetric QCD action on the noncommutative geometry is obtained by the following formula of the spectral action principle: 
\eq
I_{\textrm{SQCD}} = \textrm{Tr}_{L^2} \widetilde{\mathcal{D}}^2_0 + \left( \Psi, i \widetilde{\mathcal{D}}_0 \Psi \right),
\label{eq6.49}
\eqend
where $\widetilde{\mathcal{D}}_0$ is given by Eq.(\ref{eq6.2}).

\section{Conclusions}
In this paper we have studied how to introduce supersymmetry into the NCG models. In order to incorporate supersymmetry we enlarged the Hilbert space to include bosons as well as fermions, which constitute a supermultiplet. Then we defined the generalized Dirac operator $\mathcal{D}_M$ which operates on the supermultiplets in the Hilbert space. Although supersymmetry must be formulated in the Minkowskian signature, the axioms of NCG are stated in the Euclidean one. So, we changed the signature by the Wick rotation when we checked the axioms of NCG and calculated the spectral action. 

A vector supermultiplet was introduced as the internal fluctuation on the metric. The modified Dirac operator $\widetilde{\mathcal{D}}_M$ due to the fluctuation turned out to be supersymmetric and gauge-covariant if we introduce the internal degrees of freedom as the finite geometry. In this paper we considered the supersymmetric QCD model where the finite geometry is represented by the algebra of complex matrices. 

Following  the prescription of NCG we calculated the spectral action by using our generalized supersymmetric Dirac operator $\widetilde{\mathcal{D}}_M$. We have found that the super Yang-Mills action was successfully derived based on the spectral action principle. The other parts of the action which include kinetic terms and mass terms of quarks and their superpartners were also obtained. As a result we found that the whole supersymmetric QCD action was expressed by the simple formula of the spectral action principle.

The method proposed in this paper to incorporate supersymmetry in NCG is applicable to other models such as the supersymmetric standard model.
In this model the Higgs bosons and their superpartners are introduced as the internal fluctuation on the metric of the finite geometry. The detailed discussions and calculations on this subject will be given in a separate paper \cite{ishihara}. 
%
\bibliography{ncgsusy}
\end{document}